\documentclass[%
 amsmath,amssymb,
 aps,
preprint]{revtex4-2}

\usepackage{graphicx}
\usepackage{dcolumn}
\usepackage{bm}

\begin{document}

\title{The Large Magellanic Cloud as a source of the highest energy cosmic rays}

\author{Roger Clay}
\email{roger.clay@adelaide.edu.au}
 \affiliation{School of Physical Sciences, University of Adelaide,\\ North Terrace, Adelaide 5005, Australia}


\author{Tadeusz Wibig}
\email{tadeusz.wibig@uni.lodz.pl}
\affiliation{Faculty of Physics and Applied Informatics, University of Lodz, \\ 90-236 Lodz, Pomorska 149/153, Poland}

\begin{abstract}

The Pierre Auger Observatory has published properties of the 100 highest-energy cosmic ray events (to energies above 100 EeV) which it recorded over a 17 year period. We have examined the directional properties of these events and have taken particular note of the most energetically extreme events. We find that the most energetic events have directions which are grouped in a non-random way at the 1\% level. There is an apparent clustering in a limited region of the sky. Close to that direction is found Centaurus A, which has long been considered as a source of such particles, but we also note that a close-by dwarf galaxy, the Large Magellanic Cloud (LMC) is closer in angular terms. We examine the possibility that the LMC might be a source of observed cosmic rays at the highest energies.
\end{abstract}

\maketitle
\section{\label{sec:intro}Introduction}

Multi-messenger high-energy astrophysics now consists of a number of complementary sub-disciplines: neutrino astrophysics, X-ray astronomy, gamma-ray astronomy up to PeV energies, and cosmic ray astrophysics \cite{refId0,hess2012,lmc2015}. There are expected to be commonalities in the sources studied in those discipline areas, basically associated with the acceleration of cosmic particles and their subsequent interactions.  The photons and neutrinos travel directly to us. On the other hand, the cosmic rays are electrically charged and, when passing through cosmic magnetic fields, their paths are deflected such that, at almost every energy studied, there is no direct correlation between their observed directions and the directions directly to their sources. This means that the highest-energy cosmic ray sources and their properties have previously been largely indirectly inferred by modelling \cite{Auger_data_fits}.

It is possible that at the highest cosmic ray energies, the particle rigidities (considered as energy divided by charge) may be sufficiently high for some directional information to be retained. This had been the expectation when the present observatories for the highest energy cosmic rays (the Pierre Auger Observatory: \cite{2015172,pao2025} and the Telescope Array: \cite{COLEMAN2023102819,ABUZAYYAD201287,TOKUNO201254}) were begun. It is now clear that the composition of the nuclei making up the highest energy cosmic rays is not purely protons but is dominated by more massive (and more highly charged) nuclei (see, e.g., \cite{wibig2017universalcosmicraysenergy}) and the rigidities are less than originally expected. Furthermore, it is now known that the most energetic cosmic rays have origins outside the plane of our Galaxy and to reach us they pass through the Milky Way magnetic field \cite{Mao_2012,Unger:2015wka}, which is likely to be substantially stronger than intergalactic fields. The task for many cosmic ray studies is then to propose possible sources from the ‘zoo’ of extra-galactic objects (AGNs, starburst galaxies, galaxy clusters etc.), conjecture a cosmic ray source mechanism (popularly diffusive shock acceleration \cite{PhysRev.75.1169,Hillas_2005,Comisso_2024}) with resulting composition and particle energy spectra, propagate these particles towards an observatory through inter-galactic photon and magnetic fields, and then deflect them following their particular galactic magnetic field path. Heroic work continues in this task.

Apart from turbulent inter-galactic magnetic fields, cosmic ray particles must travel through a fog of cosmic microwave background (CMB) and infra-red photons to reach our galaxy. The former dominate numerically and result in interactions over short (astrophysical) distances of a few megaparsecs, with a resulting limited total horizon of around 100 Mpc for particles with energies above a few tens of EeV (1 EeV is 10$^{18}$ eV). This is known as the GZK cut-off \cite{PhysRevLett.16.748,1966ZhPmR...4..114Z} and it happens that both protons and more massive nuclei have rather similar attenuation in the CMB. However, even for particles with sources closer than 100 Mpc, there is still a progressive energy loss so that sources more distant than around 10 Mpc will have their particles reduced in energy before reaching us. As a result, there will be some preference in the cosmic ray flux at the highest energies for closer rather than more distant sources, both due to a normal inverse square reduction in flux with distance and a GZK attenuation of the flux of extreme energy particles from the more distant sources.

\begin{figure}[ht]
    \centering 
    \includegraphics[width=16cm]{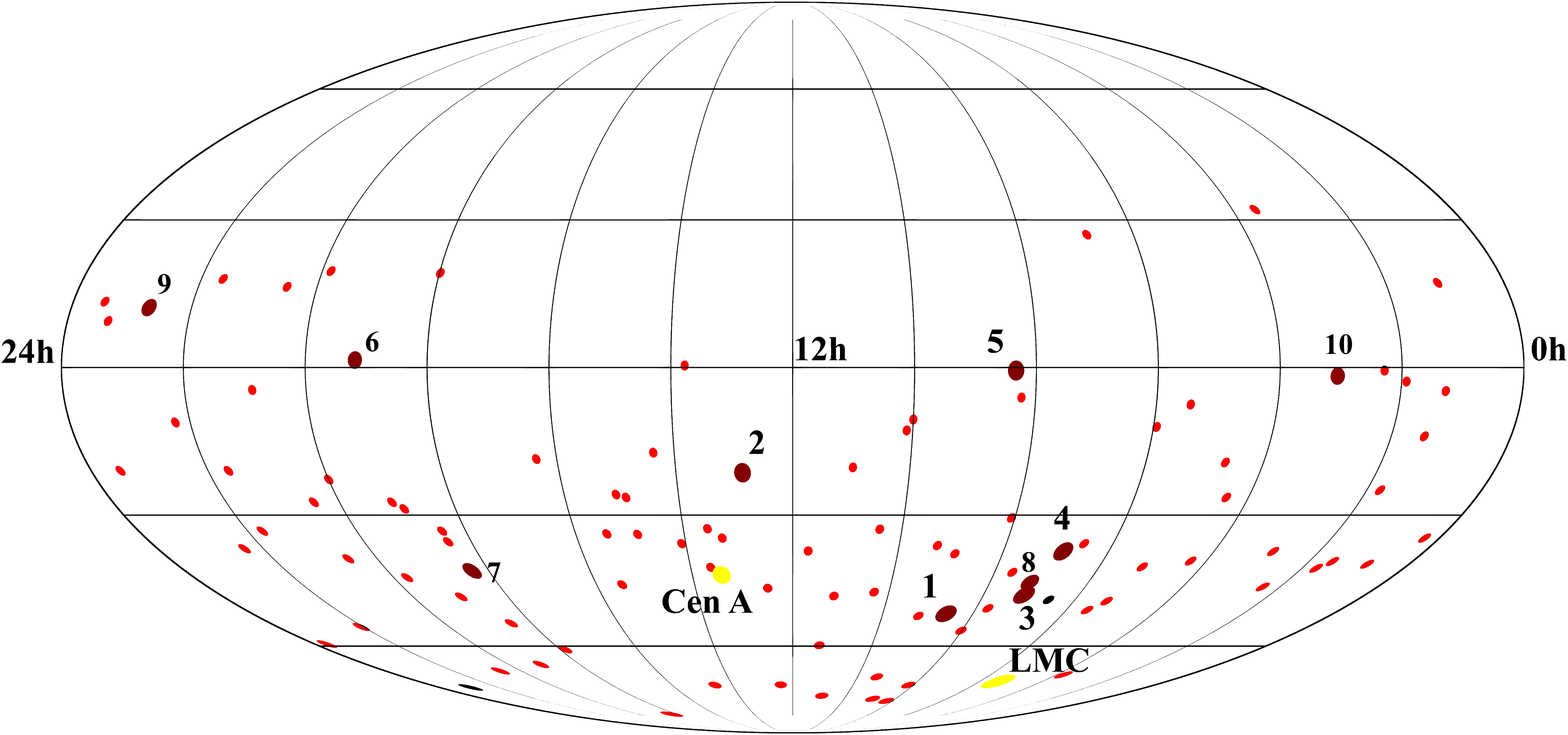}
    \caption{Sky map in celestial co-ordinates showing directions of arrival of 100 most energetic Auger events. The five most energetic events are shown as large brown spots, smaller brown spots represent next five events. Small red dots are the next 90 Auger EAS from the published list. The LMC and Cen A are shown as yellow spots. 
    \label{t234}}
\end{figure}

The Pierre Auger Collaboration (Auger), whose Observatory is located to study southerly event directions, recently published a catalog of its 100 most energetic events, recorded over a 17 year period \cite{thepierreaugercollaboration2022cataloghighestenergycosmicrays,pao2025}. This Observatory is the largest and longest running of the two prominent observatories at the highest energies, and that dataset will only increase slowly in content over the next decade, although, with a recent upgrade, the overall quality will greatly improve. If any form of positional cosmic ray astronomy is to be achieved in the short term, its data will necessarily be found using the present dataset.

We have examined the directional properties of the highest energy events in the Auger catalog and find evidence for directional clustering in those events. This apparent evidence in a skymap of the data is strong, but with a limited statistical confidence due to the limited number of available events. However, it suggests that there is a preferred source direction which may be compatible with the directions of Centaurus A (Cen A) or the Large Magellanic Cloud (LMC). Cen A, our nearest AGN, has been extensively discussed in the literature as a possible source \cite{ABREU2010314,thepierreaugercollaboration2024fluxultrahighenergycosmicrays} and, here, we will examine the LMC.  This source is somewhat closer in position to an event clustering in the highest energy events and has some properties which have not so far been greatly discussed in the context of ultra-high energy cosmic rays.

The structure of the paper is that Section \ref{frak} examines the statistical clustering properties of the Auger dataset of 100 highest energy events.  Section \ref{LMC} examines the possibility that such clustering may be due to a source in the LMC.  A brief conclusion section follows.

\section{Factorial moments analysis\label{frak}}

Searching for clustering in a dataset is currently one of the most important issues in the tasks faced by AI and applied computer science problems in general (see, e.g. \cite{clust}). The specific problems analyzed in physics are only a small part of this, but their relevance to physics studies can be extremely important. For instance, the techniques used in high-energy physics were related to the experimental search for the jet structure of hadronisation processes in high-energy particle collisions and, later, to the search for the effects of the existence of quark-gluon plasmas in collisions with very high spatial energy density.

When looking for groupings of cosmic ray directions on the celestial sphere, the obvious definition of the distance between two points ($e_1$ and $e_2$) is their angular distance $\delta(e_1,\ e_2)$, and the question of whether those points group requires an examination of whether the histogram of distances of each pair differs (significantly) from the distribution that would be observed if the points were randomly distributed on the sphere.
Using the angular diameter of a set of three, four, or more directions we can examine the higher order grouping of shower directions.

The concept of factorial cumulants was introduced to high-energy physics by
\cite{BIALAS1986703,BIALAS1988857}. 
The integral formulation of factorial moments of order $k$ defines these moments as a function of the diameter of the event groups, $\Delta \phi$ as \cite{LIPA1992300,DEWOLF19961,Kittel2000}:

\begin{equation}
    \label{factdef2}
    F_k(\Delta \phi) =  \frac{1}{ \rm norm.}\ k! 
    \sum\limits_{ \scriptscriptstyle \overrightarrow e_1<\overrightarrow e_2<...<\overrightarrow e_k} 
    \ \prod \limits_{     \renewcommand{\arraystretch}{.75}
    \begin{array}{c}
    {\scriptstyle  \rm all\ pairs} \\[-5pt] \scriptstyle m_1,\  m_2
    \end{array}
    }
    \Theta \left({\Delta\phi - \delta(\overrightarrow e_{m_1}, \overrightarrow e_{m_2})}\right)
\end{equation}
\noindent
where $\Theta$ is the Heaviside unit-step function, and the summation proceeds along all possible groups of $k$ shower directions. 

The sum gives the number of all the $k$-events observed with a diameter less than $\Delta \phi$, while the "norm." factor represents the expected number of $k$-events with a diameter less than $\Delta \phi$ that we would expect if the events were not causally related in any way.

\subsection{Results}
\begin{figure}[ht]
    \centering
    \includegraphics[width=6.6cm]{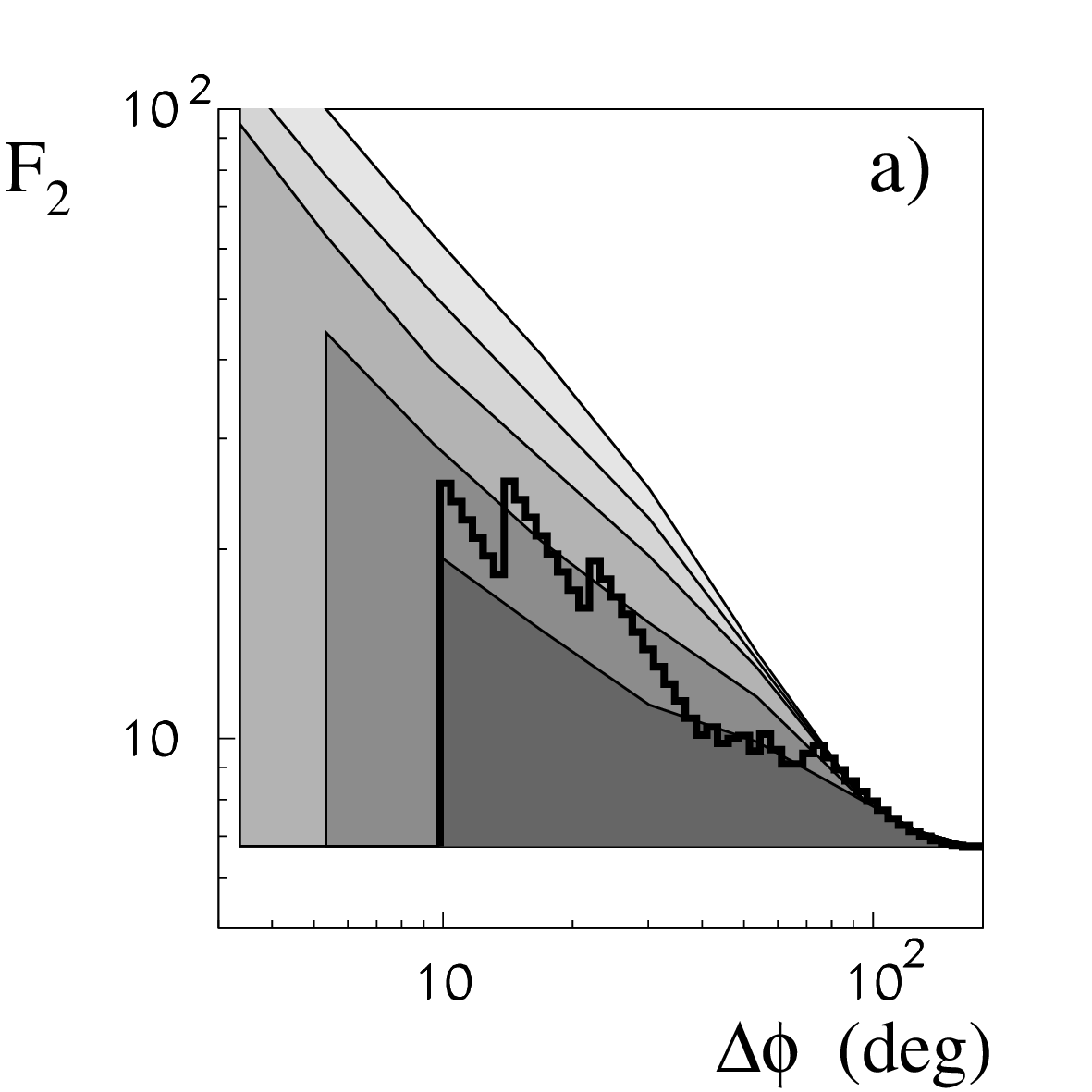}
    \includegraphics[width=6.6cm]{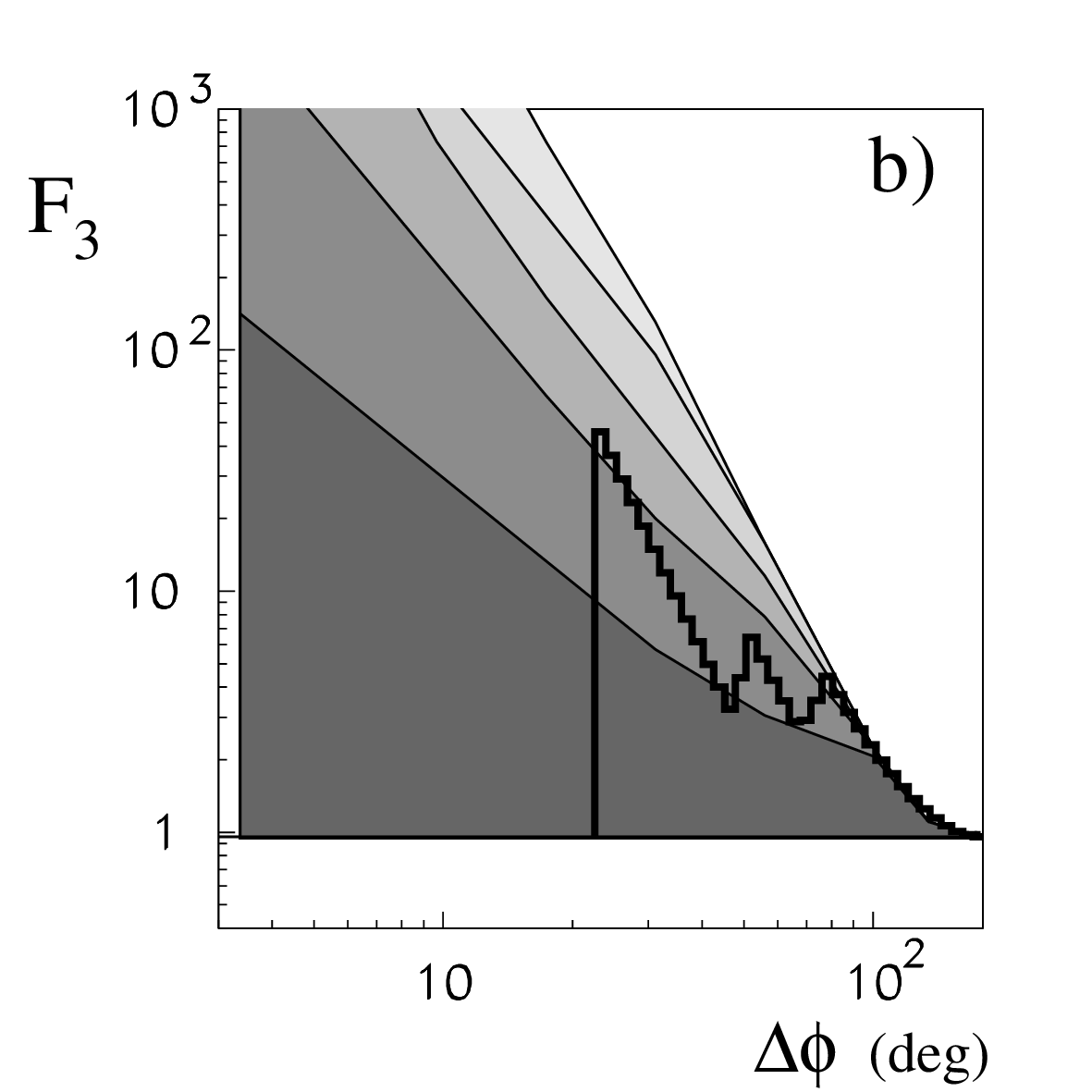}
    \includegraphics[width=6.6cm]{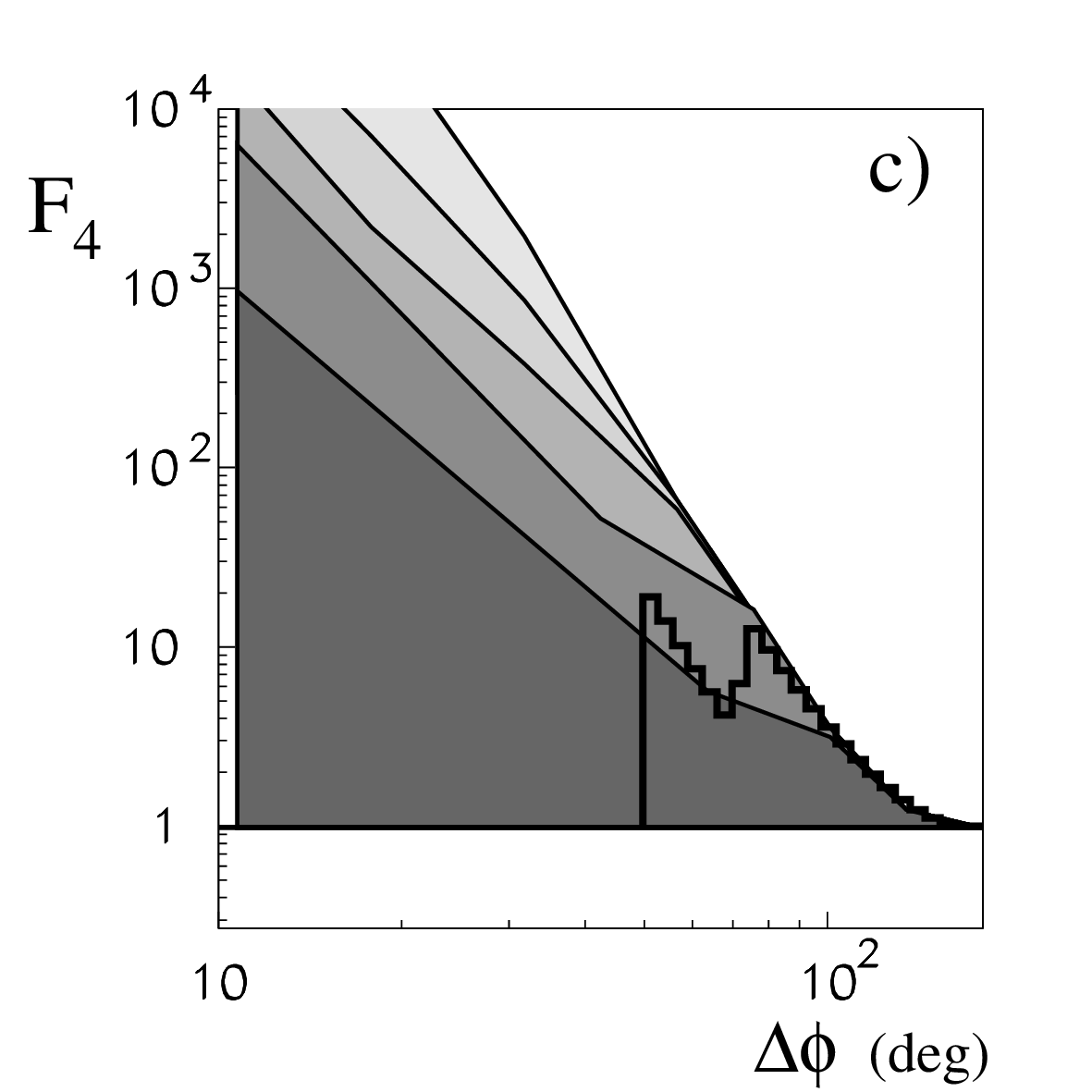}
    \caption{Factorial cumulants or second (a), third (b), fourth (c) order for the five highest-energy Auger showers as a function of cluster angular diameter. Histograms represent results obtained for the five Auger events. The shaded areas represent limits for 10\%, 1\%, 0.1\% and 0.01\% chance probabilities of the cluster of that size occurring by chance.
    \label{f2}}
\end{figure}

One particular, rather technical, problem is determining a sample of uncorrelated random events. In most, almost all, cases, such a distribution is not uniform over the available space. In collision experiments, where the production of many particles is observed and there are at least obvious constraints from conservation principles, the pool of uncorrelated particles is constructed by selecting for analysis the particles produced in different collisions. When analyzing the distribution of arrival directions of cosmic rays recorded in specific devices, such a possibility does not work. An obvious solution is to randomize the angle of Right Ascension and the choice of declination, which is defined by the acceptance of a certain apparatus for a specific energy range.

We first concentrated on the five highest-energy events observed by Auger. This ‘a priori’ decision (before plotting celestial maps) was based on the limited size of the dataset and the thought that any new physics might be found only at the very highest energies, as argued in the Introduction.
Fig. \ref{f2} shows the calculated values of factorial cumulants for the five highest-energy showers observed by Auger. As can be seen, as expected from an examination of Fig.\ref{t234}, for $F_2$ and $F_3$, the experimental histograms strongly deviate from the predicted value of 1 for the random distribution of shower directions. $F_4$, which represents the clustering of the directions of four cosmic ray showers (out of the five) now includes the angular distances of the farthest ones, and its deviation from randomness is essentially the effect of the tight grouping of events numbers 1, 3 and 4 (see Fig. \ref{t234}).

The $F_2$ and $F_3$ experimental histograms deviate to a random chance
probability of order 1\%. This value of 1\% does not appear to be strongly
statistically significant (it is far from $10^{-5}$ or $10^{-6}$ usually considered for definite discovery of a new phenomenon in experimental research) and, of course, it is thus possible (at the 1\% level) that the observed cluster of five shower events is random.

In order to see if the choice of showers is not the cause of the clustering of the selected ones, we made calculations identical to those above, discarding the five most energetic cases from the lists and taking five more for analysis. When this is done, any correlation disappears, and the cumulant values agree with the completely random distribution of directions on the celestial sphere, supporting the likelihood that, at the 1\% level, the real clustering is not random.

\begin{figure}[ht]
    \centering 
    \includegraphics[width=8cm]{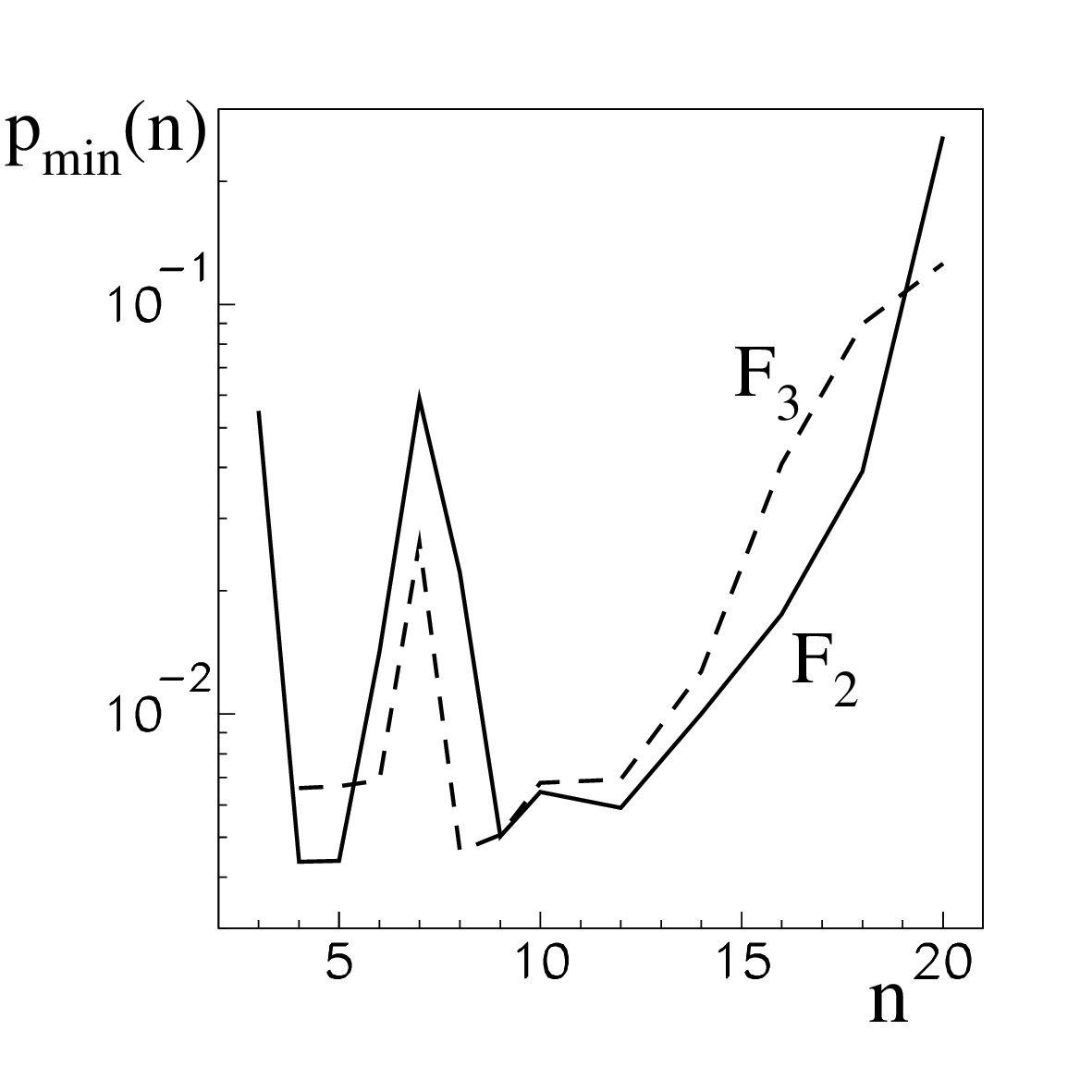}
    \caption{The probability that, among Auger $n$ observations of the most energetic shower events (counting from the highest-energy), a grouping of pairs (F2) and triples (F3) of their directions on the celestial sphere will occur by chance. It is determined as the smallest value over all group angular diameters, using plots similar to those in Fig. \ref{f2}.
    \label{pstwo}}
\end{figure}

Another test to examine the statistical significance of the observed effect is to check that it is not artificially generated by selecting \underline{exactly} the first five events from the published list of the most energetic showers recorded by Auger. We have repeated our calculations and simulations, progressively selecting other numbers of showers starting from the three with the highest-energy. Fig. \ref{pstwo} shows how the probability of clusters (pairs or threes) of directions occurring by chance changes for different numbers of selected most energetic showers. We note that the observed clustering of the first five showers is clearly not random. When four or five events are selected, there is clearly a minimum probability value of about 0.005 (below 1\%). A further increase in the number of showers selected for analysis (up to 10 -- 14) supports the conclusion of only a one per cent probability of accidental clustering of both pairs and triples of shower directions. It is worth noting that the probability of such a grouping is reduced back to $\sim$0.5\% by a shower appearing as the eighth in the Auger list (close in direction to the grouping of events 1, 3 and 4, see Fig.\ref{t234}). Of course, these particular probability values should be taken with care, their exact determination requires a more detailed analysis which, for the very limited statistics of available events, must be subject to considerable uncertainty.

\section{The Large Magellanic Cloud\label{LMC}}

As shown in Figure 1, the Large Magellanic Cloud is located in a direction close to the clustering of events 1, 3, 4, 8.  This suggests that it is important to examine this object, or a location within it, as the source of these events.  This is an unconventional speculation, but the LMC has already proven to be a significant source of energetic particles, at least up to energies of multiple-TeV \cite{fermi2010,hess2012,lmc2015}. 
At X-ray energies (to 10 keV), we now have eROSITA \cite{refId03} 
full-sky data. A useful annotated eROSITA map can be found at \cite{rosita} 
where it is relevant to note that the proximity of the LMC to us results in the LMC being the dominant extra-Milky Way source, with an appreciably stronger X-ray flux at Earth, for instance, than Centaurus A.

The LMC is not an obvious candidate for a source of particles up to energies of 100+ EeV. It is merely a dwarf galaxy satellite galaxy of our Milky Way galaxy. However, it does have properties which are thought necessary in such sources. In particular, it has a surprisingly strong, and quite extensive, magnetic field in the 30 Dorados region. Also, from our point of view, it is a very nearby source and thus suffers relatively little from diminishing flux due to an inverse square of distance, plus we note that any 100 EeV particles which might be produced there will not suffer any GZK attenuation (see, for instance, \cite{RevModPhys.71.S165} Fig. 4). The latter could mean that, at the very highest energies, an LMC particle in the GZK energy range might reach us without attenuation while a similarly energetic particle from a more distant source would inevitably suffer an energy reduction in propagation and would likely get lost in the overall cosmic ray ultra-high energy spectrum.

Within the Large Magellanic Cloud is a complex region known as 30 Doradus, or the Tarantula Nebula. Sofia measurements of this region \cite{Tram_2023} 
indicate magnetic field strengths greater than 300 $\mu$G over an angular range of $\sim$ 3 minutes of arc. At a distance of $\sim$50 kpc, this translates to a dimension of approximately 40 pc. A simple rule of thumb often used for a first examination of source regions for possible ultra-high energy (UHE) particle acceleration is a compatibility with limits found in a 
"Hillas diagram" \cite{annurev:/content/journals/10.1146/annurev.aa.22.090184.002233}.  
 The 30 Doradus source satisfies the basic Hillas criterion for heavy (iron) UHE nuclei. It does not do so for a proton composition, but, at the highest energies at which the UHE composition is measured, ‘heavy’ nuclei have been found to dominate \cite{PhysRevLett.134.021001}.

The conventional expectation for the acceleration of UHE cosmic rays is that there is diffusive shock acceleration. The LMC does have a bow shock but 30 Doradus within it does not. However, in the Sofia results we find that in 30 Doradus there is a "highly ordered yet complex B-field structure” with “possible multi expanding shells”. There is “supersonic compressive turbulence”. These properties could be important for particle acceleration but we note a caveat in that work that the B-field strength estimates do “suffer from numerous uncertainties” (discussed in the paper). Comisso, Farrar and Muzio \cite{Comisso_2024} 
have demonstrated that “particle acceleration by magnetically dominated turbulence possesses the properties needed to explain the acceleration of UHECRs”. This acceleration, through scattering off turbulence fluctuations, can be fast, with energies up to values where the Larmor radius is comparable with the coherence length of the magnetic field. In this sense, there is still a natural Hillas-like limit through the need to have some coherence being retained in trajectories within the overall magnetic field of the source.

Centaurus A is an AGN source which is also relatively close in direction to the apparent excess of events in the Pierre Auger catalog and is often thought to be a potential source of the highest energy Auger events due to an Auger anisotropy in roughly that direction \cite{2024ApJ...976...48A}. 
In passing, we note that the Auger anisotropy also adequately fits the direction shown on our event map here. However, the arrival direction of charged cosmic ray particles at Earth will not be the direction of a source as seen using uncharged messenger particles, and we must find an estimate of their directions when back-tracked to the outside of the (dominant) Milky Way magnetic field. 
 \cite{unger2025galacticmagneticfielduhecr} 
have most recently used galactic magnetic field models based on compiling the best currently available data to back-track observed directions of UHE cosmic rays to the outer limits of the Milky Way magnetic field. Their angular deflection map (their Fig. 3) shows a typical back-tracking deflection of the order of 15 degrees in the region of the Auger event cluster (rigidity $2\times 10^{19}$ eV which would be the order for a ‘heavy’ cosmic ray at the highest Auger energies). That deflection could adequately cover the LMC direction but not so easily that of Cen A. Interestingly, in that modeling, the ‘observed’ directions surrounding this direction do not have deflections pointing in consistent directions, providing the possibility that the real deflections may be quite low and certainly not greater in magnitude than the 15$^\circ$ shown by the modelling.

\section{Conclusions\label{conclu}}

The highest energy 
cosmic ray events measured by the Pierre Auger Observatory over a 17
year period show some
evidence of clustering (also compatible with a random expectation at the 1\% level), such that three out of the five most energetic events (and four of the ten highest) are found within an angular space of only about 30$^{o}$ (Fig. \ref{t234}). At those 100$+$ EeV energies, even for 'heavy' primary particles, propagation modeling indicates that any events in the general direction of that clustering should be found within 15$^{o}$ of their source direction. The only well-known source of energetic events within that limit, for the cluster described here, is the Large Magellanic Cloud, which has properties that do appear to be sufficient for the acceleration of particles to the required rigidity.

 


%

\bibliography{rclay}{}

\providecommand{\noopsort}[1]{}\providecommand{\singleletter}[1]{#1}%
\begin{thebibliography}{35}%
\makeatletter
\providecommand \@ifxundefined [1]{%
 \@ifx{#1\undefined}
}%
\providecommand \@ifnum [1]{%
 \ifnum #1\expandafter \@firstoftwo
 \else \expandafter \@secondoftwo
 \fi
}%
\providecommand \@ifx [1]{%
 \ifx #1\expandafter \@firstoftwo
 \else \expandafter \@secondoftwo
 \fi
}%
\providecommand \natexlab [1]{#1}%
\providecommand \enquote  [1]{``#1''}%
\providecommand \bibnamefont  [1]{#1}%
\providecommand \bibfnamefont [1]{#1}%
\providecommand \citenamefont [1]{#1}%
\providecommand \href@noop [0]{\@secondoftwo}%
\providecommand \href [0]{\begingroup \@sanitize@url \@href}%
\providecommand \@href[1]{\@@startlink{#1}\@@href}%
\providecommand \@@href[1]{\endgroup#1\@@endlink}%
\providecommand \@sanitize@url [0]{\catcode `\\12\catcode `\$12\catcode `\&12\catcode `\#12\catcode `\^12\catcode `\_12\catcode `\%12\relax}%
\providecommand \@@startlink[1]{}%
\providecommand \@@endlink[0]{}%
\providecommand \url  [0]{\begingroup\@sanitize@url \@url }%
\providecommand \@url [1]{\endgroup\@href {#1}{\urlprefix }}%
\providecommand \urlprefix  [0]{URL }%
\providecommand \Eprint [0]{\href }%
\providecommand \doibase [0]{https://doi.org/}%
\providecommand \selectlanguage [0]{\@gobble}%
\providecommand \bibinfo  [0]{\@secondoftwo}%
\providecommand \bibfield  [0]{\@secondoftwo}%
\providecommand \translation [1]{[#1]}%
\providecommand \BibitemOpen [0]{}%
\providecommand \bibitemStop [0]{}%
\providecommand \bibitemNoStop [0]{.\EOS\space}%
\providecommand \EOS [0]{\spacefactor3000\relax}%
\providecommand \BibitemShut  [1]{\csname bibitem#1\endcsname}%
\let\auto@bib@innerbib\@empty
\bibitem [{\citenamefont {\mbox{ K. W. Wo\'{z}niak (for the \uppercase{CREDO C}ollaboration)}}(2019)}]{refId0}%
  \BibitemOpen
  \bibfield  {author} {\bibinfo {author} {\bibnamefont {\mbox{ K. W. Wo\'{z}niak (for the \uppercase{CREDO C}ollaboration)}}},\ }\bibfield  {title} {\bibinfo {title} {Detection of cosmic-ray ensembles with \uppercase{CREDO}},\ }\href@noop {} {\bibfield  {journal} {\bibinfo  {journal} {EPJ Web Conf.}\ }\textbf {\bibinfo {volume} {208}},\ \bibinfo {pages} {15006} (\bibinfo {year} {2019})}\BibitemShut {NoStop}%
\bibitem [{\citenamefont {\mbox{A. Abramowski et al. (H.E.S.S. Collaboration)}}(2012)}]{hess2012}%
  \BibitemOpen
  \bibfield  {author} {\bibinfo {author} {\bibnamefont {\mbox{A. Abramowski et al. (H.E.S.S. Collaboration)}}},\ }\bibfield  {title} {\bibinfo {title} {Discovery of $\gamma$-ray emission from the extragalactic pulsar wind nebula \uppercase{N 157B} with \uppercase{H.E.S.S.}},\ }\href {https://doi.org/10.1051/0004-6361/201219906} {\bibfield  {journal} {\bibinfo  {journal} {A\&A}\ }\textbf {\bibinfo {volume} {545}},\ \bibinfo {pages} {L2} (\bibinfo {year} {2012})}\BibitemShut {NoStop}%
\bibitem [{\citenamefont {\mbox{A. Abramowski et al. (H.E.S.S. Collaboration)}}(2015)}]{lmc2015}%
  \BibitemOpen
  \bibfield  {author} {\bibinfo {author} {\bibnamefont {\mbox{A. Abramowski et al. (H.E.S.S. Collaboration)}}},\ }\bibfield  {title} {\bibinfo {title} {The exceptionally powerful \uppercase{T}e\uppercase{V} $\gamma$-ray emitters in the \uppercase{L}arge \uppercase{M}agellanic \uppercase{C}loud},\ }\href {https://doi.org/10.1126/science.1261313} {\bibfield  {journal} {\bibinfo  {journal} {Science}\ }\textbf {\bibinfo {volume} {347}},\ \bibinfo {pages} {406} (\bibinfo {year} {2015})}\BibitemShut {NoStop}%
\bibitem [{\citenamefont {\mbox{A. A. Halim et al. (Pierre Auger Collaboration)}}(2024)}]{Auger_data_fits}%
  \BibitemOpen
  \bibfield  {author} {\bibinfo {author} {\bibnamefont {\mbox{A. A. Halim et al. (Pierre Auger Collaboration)}}},\ }\bibfield  {title} {\bibinfo {title} {Constraining models for the origin of ultra-high-energy cosmic rays with a novel combined analysis of arrival directions, spectrum, and composition data measured at the \mbox{Pierre Auger Observatory}},\ }\href {https://doi.org/10.1088/1475-7516/2024/01/022} {\bibfield  {journal} {\bibinfo  {journal} {Journal of Cosmology and Astroparticle Physics}\ }\textbf {\bibinfo {volume} {2024}}\bibinfo  {number} { (01)},\ \bibinfo {pages} {022}}\BibitemShut {NoStop}%
\bibitem [{\citenamefont {\mbox{A. Aab et al. (Pierre Auger Collaboration)}}(2015)}]{2015172}%
  \BibitemOpen
\bibfield  {number} {  }\bibfield  {author} {\bibinfo {author} {\bibnamefont {\mbox{A. Aab et al. (Pierre Auger Collaboration)}}},\ }\bibfield  {title} {\bibinfo {title} {\mbox{The Pierre Auger} \uppercase{C}osmic \uppercase{R}ay \uppercase{O}bservatory},\ }\href {https://doi.org/https://doi.org/10.1016/j.nima.2015.06.058} {\bibfield  {journal} {\bibinfo  {journal} {Nuclear Instruments and Methods in Physics Research A}\ }\textbf {\bibinfo {volume} {798}},\ \bibinfo {pages} {172} (\bibinfo {year} {2015})}\BibitemShut {NoStop}%
\bibitem [{\citenamefont {\mbox{A. A. Halim et al. (\uppercase{P}ierre \uppercase{A}uger \uppercase{C}ollaboration)}}(2025)}]{pao2025}%
  \BibitemOpen
  \bibfield  {author} {\bibinfo {author} {\bibnamefont {\mbox{A. A. Halim et al. (\uppercase{P}ierre \uppercase{A}uger \uppercase{C}ollaboration)}}},\ }\bibfield  {title} {\bibinfo {title} {The \uppercase{P}ierre \uppercase{A}uger \uppercase{O}bservatory open data},\ }\href {https://doi.org/https://doi.org/10.1140/epjc/s10052-024-13560-5} {\bibfield  {journal} {\bibinfo  {journal} {Eur. Phys. J. C}\ }\textbf {\bibinfo {volume} {85}},\ \bibinfo {pages} {70} (\bibinfo {year} {2025})}\BibitemShut {NoStop}%
\bibitem [{\citenamefont {\mbox{A. Coleman et al. (TA Collaboration)}}(2023)}]{COLEMAN2023102819}%
  \BibitemOpen
  \bibfield  {author} {\bibinfo {author} {\bibnamefont {\mbox{A. Coleman et al. (TA Collaboration)}}},\ }\bibfield  {title} {\bibinfo {title} {Ultra high energy cosmic rays the intersection of the cosmic and energy frontiers},\ }\href {https://doi.org/https://doi.org/10.1016/j.astropartphys.2023.102819} {\bibfield  {journal} {\bibinfo  {journal} {Astroparticle Physics}\ }\textbf {\bibinfo {volume} {149}},\ \bibinfo {pages} {102819} (\bibinfo {year} {2023})}\BibitemShut {NoStop}%
\bibitem [{\citenamefont {\mbox{Abu-Zayyad, T., et al.}}(2012)}]{ABUZAYYAD201287}%
  \BibitemOpen
  \bibfield  {author} {\bibinfo {author} {\bibnamefont {\mbox{Abu-Zayyad, T., et al.}}},\ }\bibfield  {title} {\bibinfo {title} {The surface detector array of the telescope array experiment},\ }\href {https://doi.org/https://doi.org/10.1016/j.nima.2012.05.079} {\bibfield  {journal} {\bibinfo  {journal} {Nuclear Instruments and Methods in Physics Research A}\ }\textbf {\bibinfo {volume} {689}},\ \bibinfo {pages} {87} (\bibinfo {year} {2012})}\BibitemShut {NoStop}%
\bibitem [{\citenamefont {\mbox{Tokuno, H. et al.}}(2012)}]{TOKUNO201254}%
  \BibitemOpen
  \bibfield  {author} {\bibinfo {author} {\bibnamefont {\mbox{Tokuno, H. et al.}}},\ }\bibfield  {title} {\bibinfo {title} {New air fluorescence detectors employed in the telescope array experiment},\ }\href {https://doi.org/https://doi.org/10.1016/j.nima.2012.02.044} {\bibfield  {journal} {\bibinfo  {journal} {Nuclear Instruments and Methods in Physics Research A}\ }\textbf {\bibinfo {volume} {676}},\ \bibinfo {pages} {54} (\bibinfo {year} {2012})}\BibitemShut {NoStop}%
\bibitem [{\citenamefont {Wibig}\ and\ \citenamefont {Wolfendale}(2017)}]{wibig2017universalcosmicraysenergy}%
  \BibitemOpen
  \bibfield  {author} {\bibinfo {author} {\bibfnamefont {T.}~\bibnamefont {Wibig}}\ and\ \bibinfo {author} {\bibfnamefont {A.~W.}\ \bibnamefont {Wolfendale}},\ }\href {https://arxiv.org/abs/1706.05872} {\bibinfo {title} {Universal cosmic rays energy spectrum and the mass composition at the 'ankle' and above}} (\bibinfo {year} {2017}),\ \Eprint {https://arxiv.org/abs/1706.05872} {arXiv:1706.05872 [astro-ph.HE]} \BibitemShut {NoStop}%
\bibitem [{\citenamefont {\mbox{ S. A. Mao et al.}}(2012)}]{Mao_2012}%
  \BibitemOpen
  \bibfield  {author} {\bibinfo {author} {\bibnamefont {\mbox{ S. A. Mao et al.}}},\ }\bibfield  {title} {\bibinfo {title} {Magnetic field structure of the \uppercase{L}arge \uppercase{M}agellanic \uppercase{C}loud from \uppercase{F}araday rotation measures of diffuse polarized emission},\ }\href {https://doi.org/10.1088/0004-637X/759/1/25} {\bibfield  {journal} {\bibinfo  {journal} {The Astrophysical Journal}\ }\textbf {\bibinfo {volume} {759}},\ \bibinfo {pages} {25} (\bibinfo {year} {2012})}\BibitemShut {NoStop}%
\bibitem [{\citenamefont {Unger}\ and\ \citenamefont {Farrar}(2015)}]{Unger:2015wka}%
  \BibitemOpen
  \bibfield  {author} {\bibinfo {author} {\bibfnamefont {M.}~\bibnamefont {Unger}}\ and\ \bibinfo {author} {\bibfnamefont {G.}~\bibnamefont {Farrar}},\ }\href@noop {} {\bibinfo {title} {{(In)Feasability of Studying Ultra-High-Energy Cosmic Rays with Smartphones}}} (\bibinfo {year} {2015})\BibitemShut {NoStop}%
\bibitem [{\citenamefont {Fermi}(1949)}]{PhysRev.75.1169}%
  \BibitemOpen
  \bibfield  {author} {\bibinfo {author} {\bibfnamefont {E.}~\bibnamefont {Fermi}},\ }\bibfield  {title} {\bibinfo {title} {On the origin of the cosmic radiation},\ }\href {https://doi.org/10.1103/PhysRev.75.1169} {\bibfield  {journal} {\bibinfo  {journal} {Phys. Rev.}\ }\textbf {\bibinfo {volume} {75}},\ \bibinfo {pages} {1169} (\bibinfo {year} {1949})}\BibitemShut {NoStop}%
\bibitem [{\citenamefont {Hillas}(2005)}]{Hillas_2005}%
  \BibitemOpen
  \bibfield  {author} {\bibinfo {author} {\bibfnamefont {A.~M.}\ \bibnamefont {Hillas}},\ }\bibfield  {title} {\bibinfo {title} {Can diffusive shock acceleration in supernova remnants account for high-energy galactic cosmic rays?},\ }\href {https://doi.org/10.1088/0954-3899/31/5/R02} {\bibfield  {journal} {\bibinfo  {journal} {Journal of Physics G: Nuclear and Particle Physics}\ }\textbf {\bibinfo {volume} {31}},\ \bibinfo {pages} {R95} (\bibinfo {year} {2005})}\BibitemShut {NoStop}%
\bibitem [{\citenamefont {Comisso}\ \emph {et~al.}(2024)\citenamefont {Comisso}, \citenamefont {Farrar},\ and\ \citenamefont {Muzio}}]{Comisso_2024}%
  \BibitemOpen
  \bibfield  {author} {\bibinfo {author} {\bibfnamefont {L.}~\bibnamefont {Comisso}}, \bibinfo {author} {\bibfnamefont {G.~R.}\ \bibnamefont {Farrar}},\ and\ \bibinfo {author} {\bibfnamefont {M.~S.}\ \bibnamefont {Muzio}},\ }\bibfield  {title} {\bibinfo {title} {Ultra-high-energy cosmic rays accelerated by magnetically dominated turbulence},\ }\href {https://doi.org/10.3847/2041-8213/ad955f} {\bibfield  {journal} {\bibinfo  {journal} {The Astrophysical Journal Letters}\ }\textbf {\bibinfo {volume} {977}},\ \bibinfo {pages} {L18} (\bibinfo {year} {2024})}\BibitemShut {NoStop}%
\bibitem [{\citenamefont {Greisen}(1966)}]{PhysRevLett.16.748}%
  \BibitemOpen
  \bibfield  {author} {\bibinfo {author} {\bibfnamefont {K.}~\bibnamefont {Greisen}},\ }\bibfield  {title} {\bibinfo {title} {End to the cosmic-ray spectrum?},\ }\href {https://doi.org/10.1103/PhysRevLett.16.748} {\bibfield  {journal} {\bibinfo  {journal} {Phys. Rev. Lett.}\ }\textbf {\bibinfo {volume} {16}},\ \bibinfo {pages} {748} (\bibinfo {year} {1966})}\BibitemShut {NoStop}%
\bibitem [{\citenamefont {{Zatsepin}}\ and\ \citenamefont {{Kuz'min}}(1966)}]{1966ZhPmR...4..114Z}%
  \BibitemOpen
  \bibfield  {author} {\bibinfo {author} {\bibfnamefont {G.~T.}\ \bibnamefont {{Zatsepin}}}\ and\ \bibinfo {author} {\bibfnamefont {V.~A.}\ \bibnamefont {{Kuz'min}}},\ }\bibfield  {title} {\bibinfo {title} {{Upper Limit of the Spectrum of Cosmic Rays}},\ }\href@noop {} {\bibfield  {journal} {\bibinfo  {journal} {ZhETF Pisma Redaktsiiu}\ }\textbf {\bibinfo {volume} {4}},\ \bibinfo {pages} {114} (\bibinfo {year} {1966})}\BibitemShut {NoStop}%
\bibitem [{\citenamefont {\mbox{A. A. Halim et al. (The Pierre Auger Collaboration)}}(2022)}]{thepierreaugercollaboration2022cataloghighestenergycosmicrays}%
  \BibitemOpen
  \bibfield  {author} {\bibinfo {author} {\bibnamefont {\mbox{A. A. Halim et al. (The Pierre Auger Collaboration)}}},\ }\href {https://arxiv.org/abs/2211.16020} {\bibinfo {title} {A catalog of the highest-energy cosmic rays recorded during \uppercase{P}hase \uppercase{I} of operation of the \uppercase{P}ierre \uppercase{A}uger \uppercase{O}bservatory}} (\bibinfo {year} {2022}),\ \Eprint {https://arxiv.org/abs/2211.16020} {arXiv:2211.16020 [astro-ph.HE]} \BibitemShut {NoStop}%
\bibitem [{\citenamefont {\mbox{ P. Abreu et al. (\uppercase{P}ierre \uppercase{A}uger \uppercase{C}ollaboration)}}(2010)}]{ABREU2010314}%
  \BibitemOpen
  \bibfield  {author} {\bibinfo {author} {\bibnamefont {\mbox{ P. Abreu et al. (\uppercase{P}ierre \uppercase{A}uger \uppercase{C}ollaboration)}}},\ }\bibfield  {title} {\bibinfo {title} {Update on the correlation of the highest energy cosmic rays with nearby extragalactic matter},\ }\href {https://doi.org/https://doi.org/10.1016/j.astropartphys.2010.08.010} {\bibfield  {journal} {\bibinfo  {journal} {Astroparticle Physics}\ }\textbf {\bibinfo {volume} {34}},\ \bibinfo {pages} {314} (\bibinfo {year} {2010})}\BibitemShut {NoStop}%
\bibitem [{\citenamefont {\mbox{A. A. Halim et al.} (The \uppercase{P}ierre~\uppercase{A}uger \uppercase{C}ollaboration)}(2024)}]{thepierreaugercollaboration2024fluxultrahighenergycosmicrays}%
  \BibitemOpen
  \bibfield  {author} {\bibinfo {author} {\bibnamefont {\mbox{A. A. Halim et al.} (The \uppercase{P}ierre~\uppercase{A}uger \uppercase{C}ollaboration)}},\ }\href {https://arxiv.org/abs/2407.06874} {\bibinfo {title} {The flux of ultra-high-energy cosmic rays along the supergalactic plane measured at the \uppercase{P}ierre \uppercase{A}uger \uppercase{P}bservatory}} (\bibinfo {year} {2024}),\ \Eprint {https://arxiv.org/abs/2407.06874} {arXiv:2407.06874 [astro-ph.HE]} \BibitemShut {NoStop}%
\bibitem [{\citenamefont {Alasalı}\ and\ \citenamefont {Ortakcı}(2024)}]{clust}%
  \BibitemOpen
  \bibfield  {author} {\bibinfo {author} {\bibfnamefont {T.}~\bibnamefont {Alasalı}}\ and\ \bibinfo {author} {\bibfnamefont {Y.}~\bibnamefont {Ortakcı}},\ }\bibfield  {title} {\bibinfo {title} {Clustering techniques in data mining: A survey of methods challenges, and applications},\ }\href {https://doi.org/https://doi.org/10.53070/bbd.1421527} {\bibfield  {journal} {\bibinfo  {journal} {Computer Science}\ }\textbf {\bibinfo {volume} {9}},\ \bibinfo {pages} {32} (\bibinfo {year} {2024})}\BibitemShut {NoStop}%
\bibitem [{\citenamefont {Bialas}\ and\ \citenamefont {Peschanski}(1986)}]{BIALAS1986703}%
  \BibitemOpen
  \bibfield  {author} {\bibinfo {author} {\bibfnamefont {A.}~\bibnamefont {Bialas}}\ and\ \bibinfo {author} {\bibfnamefont {R.}~\bibnamefont {Peschanski}},\ }\bibfield  {title} {\bibinfo {title} {Moments of rapidity distributions as a measure of short-range fluctuations in high-energy collisions},\ }\href {https://doi.org/https://doi.org/10.1016/0550-3213(86)90386-X} {\bibfield  {journal} {\bibinfo  {journal} {Nuclear Physics B}\ }\textbf {\bibinfo {volume} {273}},\ \bibinfo {pages} {703} (\bibinfo {year} {1986})}\BibitemShut {NoStop}%
\bibitem [{\citenamefont {Bialas}\ and\ \citenamefont {Peschanski}(1988)}]{BIALAS1988857}%
  \BibitemOpen
  \bibfield  {author} {\bibinfo {author} {\bibfnamefont {A.}~\bibnamefont {Bialas}}\ and\ \bibinfo {author} {\bibfnamefont {R.}~\bibnamefont {Peschanski}},\ }\bibfield  {title} {\bibinfo {title} {Intermittency in multiparticle production at high energy},\ }\href {https://doi.org/https://doi.org/10.1016/0550-3213(88)90131-9} {\bibfield  {journal} {\bibinfo  {journal} {Nuclear Physics B}\ }\textbf {\bibinfo {volume} {308}},\ \bibinfo {pages} {857} (\bibinfo {year} {1988})}\BibitemShut {NoStop}%
\bibitem [{\citenamefont {Lipa}\ \emph {et~al.}(1992)\citenamefont {Lipa}, \citenamefont {Carruthers}, \citenamefont {Eggers},\ and\ \citenamefont {Buschbeck}}]{LIPA1992300}%
  \BibitemOpen
  \bibfield  {author} {\bibinfo {author} {\bibfnamefont {P.}~\bibnamefont {Lipa}}, \bibinfo {author} {\bibfnamefont {P.}~\bibnamefont {Carruthers}}, \bibinfo {author} {\bibfnamefont {H.}~\bibnamefont {Eggers}},\ and\ \bibinfo {author} {\bibfnamefont {B.}~\bibnamefont {Buschbeck}},\ }\bibfield  {title} {\bibinfo {title} {The correlation integral as probe of multiparticle correlations},\ }\href {https://doi.org/https://doi.org/10.1016/0370-2693(92)91468-O} {\bibfield  {journal} {\bibinfo  {journal} {Physics Letters B}\ }\textbf {\bibinfo {volume} {285}},\ \bibinfo {pages} {300} (\bibinfo {year} {1992})}\BibitemShut {NoStop}%
\bibitem [{\citenamefont {{De Wolf}}\ \emph {et~al.}(1996)\citenamefont {{De Wolf}}, \citenamefont {Dremin},\ and\ \citenamefont {Kittel}}]{DEWOLF19961}%
  \BibitemOpen
  \bibfield  {author} {\bibinfo {author} {\bibfnamefont {E.}~\bibnamefont {{De Wolf}}}, \bibinfo {author} {\bibfnamefont {I.}~\bibnamefont {Dremin}},\ and\ \bibinfo {author} {\bibfnamefont {W.}~\bibnamefont {Kittel}},\ }\bibfield  {title} {\bibinfo {title} {Scaling laws for density correlations and fluctuations in multiparticle dynamics},\ }\href {https://doi.org/https://doi.org/10.1016/0370-1573(95)00069-0} {\bibfield  {journal} {\bibinfo  {journal} {Physics Reports}\ }\textbf {\bibinfo {volume} {270}},\ \bibinfo {pages} {1} (\bibinfo {year} {1996})}\BibitemShut {NoStop}%
\bibitem [{\citenamefont {Kittel}(2000)}]{Kittel2000}%
  \BibitemOpen
  \bibfield  {author} {\bibinfo {author} {\bibfnamefont {W.}~\bibnamefont {Kittel}},\ }\bibinfo {title} {Correlations and fluctuations in high-energy collisions},\ in\ \href {https://doi.org/10.1007/978-94-011-4126-0\_6} {\emph {\bibinfo {booktitle} {Particle Production Spanning MeV and TeV Energies}}}\ (\bibinfo  {publisher} {Springer Netherlands},\ \bibinfo {address} {Dordrecht},\ \bibinfo {year} {2000})\ Chap.~\bibinfo {chapter} {6}, pp.\ \bibinfo {pages} {157--182}\BibitemShut {NoStop}%
\bibitem [{\citenamefont {\mbox{A. A. Abdo et al.}}(2010)}]{fermi2010}%
  \BibitemOpen
  \bibfield  {author} {\bibinfo {author} {\bibnamefont {\mbox{A. A. Abdo et al.}}},\ }\bibfield  {title} {\bibinfo {title} {Observations of the \uppercase{L}arge \uppercase{M}agellanic \uppercase{C}loud with \uppercase{F}ermi},\ }\href {https://doi.org/10.1051/0004-6361/200913474} {\bibfield  {journal} {\bibinfo  {journal} {A\&A}\ }\textbf {\bibinfo {volume} {512}},\ \bibinfo {pages} {A7} (\bibinfo {year} {2010})}\BibitemShut {NoStop}%
\bibitem [{\citenamefont {{Merloni, A.}}\ and\ \citenamefont {et~al.}(2024)}]{refId03}%
  \BibitemOpen
  \bibfield  {author} {\bibinfo {author} {\bibnamefont {{Merloni, A.}}}\ and\ \bibinfo {author} {\bibnamefont {et~al.}},\ }\bibfield  {title} {\bibinfo {title} {The \uppercase{SRG}/e\uppercase{ROSITA} all-sky survey - first x-ray catalogues and data release of the western galactic hemisphere},\ }\href {https://doi.org/10.1051/0004-6361/202347165} {\bibfield  {journal} {\bibinfo  {journal} {A\&A}\ }\textbf {\bibinfo {volume} {682}},\ \bibinfo {pages} {A34} (\bibinfo {year} {2024})}\BibitemShut {NoStop}%
\bibitem [{\citenamefont {\mbox{Sky \& Telescope}}(2020)}]{rosita}%
  \BibitemOpen
  \bibfield  {author} {\bibinfo {author} {\bibnamefont {\mbox{Sky \& Telescope}}},\ }\bibfield  {title} {\bibinfo {title} {\mbox{The first all-sky \uppercase{X}-ray map to be} released in 30 years reveals new wonders of the hot and energetic universe},\ }\href@noop {} {\bibfield  {journal} {\bibinfo  {journal} {https://skyandtelescope.org/astronomy-news/first-all-sky-map-erosita}\ } (\bibinfo {year} {2020})}\BibitemShut {NoStop}%
\bibitem [{\citenamefont {Cronin}(1999)}]{RevModPhys.71.S165}%
  \BibitemOpen
  \bibfield  {author} {\bibinfo {author} {\bibfnamefont {J.~W.}\ \bibnamefont {Cronin}},\ }\bibfield  {title} {\bibinfo {title} {Cosmic rays: the most energetic particles in the universe},\ }\href {https://doi.org/10.1103/RevModPhys.71.S165} {\bibfield  {journal} {\bibinfo  {journal} {Rev. Mod. Phys.}\ }\textbf {\bibinfo {volume} {71}},\ \bibinfo {pages} {S165} (\bibinfo {year} {1999})}\BibitemShut {NoStop}%
\bibitem [{\citenamefont {\mbox{ L. N. Tram, et al.}}(2023)}]{Tram_2023}%
  \BibitemOpen
  \bibfield  {author} {\bibinfo {author} {\bibnamefont {\mbox{ L. N. Tram, et al.}}},\ }\bibfield  {title} {\bibinfo {title} {\uppercase{SOFIA} observations of 30 \uppercase{D}oradus. \uppercase{II}. \uppercase{M}agnetic fields and large-scale gas kinematics},\ }\href {https://doi.org/10.3847/1538-4357/acaab0} {\bibfield  {journal} {\bibinfo  {journal} {The Astrophysical Journal}\ }\textbf {\bibinfo {volume} {946}},\ \bibinfo {pages} {8} (\bibinfo {year} {2023})}\BibitemShut {NoStop}%
\bibitem [{\citenamefont {Hillas}(1984)}]{annurev:/content/journals/10.1146/annurev.aa.22.090184.002233}%
  \BibitemOpen
  \bibfield  {author} {\bibinfo {author} {\bibfnamefont {A.~M.}\ \bibnamefont {Hillas}},\ }\bibfield  {title} {\bibinfo {title} {The origin of ultra-high-energy cosmic rays},\ }\href {https://doi.org/https://doi.org/10.1146/annurev.aa.22.090184.002233} {\bibfield  {journal} {\bibinfo  {journal} {Annual Review of Astronomy and Astrophysics}\ }\textbf {\bibinfo {volume} {22}},\ \bibinfo {pages} {425} (\bibinfo {year} {1984})}\BibitemShut {NoStop}%
\bibitem [{\citenamefont {\mbox{A. A. Halim et al. (The Pierre Auger Collaboration)}}(2025)}]{PhysRevLett.134.021001}%
  \BibitemOpen
  \bibfield  {author} {\bibinfo {author} {\bibnamefont {\mbox{A. A. Halim et al. (The Pierre Auger Collaboration)}}},\ }\bibfield  {title} {\bibinfo {title} {Inference of the mass composition of cosmic rays with energies from ${10}^{18.5}$ to ${10}^{20}\text{ }\mathrm{eV}$ using the \uppercase{P}ierre \uppercase{A}uger \uppercase{O}bservatory and deep learning},\ }\href {https://doi.org/10.1103/PhysRevLett.134.021001} {\bibfield  {journal} {\bibinfo  {journal} {Phys. Rev. Lett.}\ }\textbf {\bibinfo {volume} {134}},\ \bibinfo {pages} {021001} (\bibinfo {year} {2025})}\BibitemShut {NoStop}%
\bibitem [{\citenamefont {\mbox{A. A. Halim et al. (The Pierre Auger Collaboration)}}(2024)}]{2024ApJ...976...48A}%
  \BibitemOpen
  \bibfield  {author} {\bibinfo {author} {\bibnamefont {\mbox{A. A. Halim et al. (The Pierre Auger Collaboration)}}},\ }\bibfield  {title} {\bibinfo {title} {{Large-scale Cosmic-ray Anisotropies with 19 yr of Data from the Pierre Auger Observatory}},\ }\href {https://doi.org/10.3847/1538-4357/ad843b} {\bibfield  {journal} {\bibinfo  {journal} {The Astrophysical Journal}\ }\textbf {\bibinfo {volume} {976}},\ \bibinfo {eid} {48} (\bibinfo {year} {2024})},\ \Eprint {https://arxiv.org/abs/2408.05292} {arXiv:2408.05292 [astro-ph.HE]} \BibitemShut {NoStop}%
\bibitem [{\citenamefont {Unger}\ and\ \citenamefont {Farrar}(2025)}]{unger2025galacticmagneticfielduhecr}%
  \BibitemOpen
  \bibfield  {author} {\bibinfo {author} {\bibfnamefont {M.}~\bibnamefont {Unger}}\ and\ \bibinfo {author} {\bibfnamefont {G.~R.}\ \bibnamefont {Farrar}},\ }\href {https://arxiv.org/abs/2502.15876} {\bibinfo {title} {The \uppercase{G}alactic magnetic field and \uppercase{UHECR} deflections}} (\bibinfo {year} {2025}),\ \Eprint {https://arxiv.org/abs/2502.15876} {arXiv:2502.15876 [astro-ph.HE]} \BibitemShut {NoStop}%
\end{thebibliography}%



\end{document}